\documentstyle[prl,floats,aps]{revtex}
\newcommand{\be}{\begin{equation}}
\newcommand{\ee}{\end{equation}}
\begin{document} 
\draft    
\renewcommand{\topfraction}{0.99}
\renewcommand{\bottomfraction}{0.99}
\twocolumn[\hsize\textwidth\columnwidth\hsize\csname 
@twocolumnfalse\endcsname

\title{On The Varying Speed of Light in a Brane-Induced FRW Universe} 
\author{Stephon H.S. Alexander}
\address{~\\
 Physics Department, Brown University, Providence, RI 02912, USA
\\Blackett Laboratory, Imperial College, Prince Consort Road, London SW7
2BZ, U.K.}

\maketitle

\begin{abstract}
We investigate a string/M theoretic realization of the varying speed of light scenario. We consider a 3+1 dimensional probe-brane universe in the background of a black hole in the bulk formed by a stack of branes, in the spirit of Kiritsis (hep-th/9906206). We generalize the dynamics of the system at hand by including rotation and Hubble damping of the bulk space-time and show that this 
may lead to a mechanism to stabilize the brane-universe and hence fix the speed of light at late times.    
\end{abstract}

]

\section{Introduction and Motivation}

One of the main challenges for modern cosmology is the Horizon Problem: the
cosmic microwave background is observed to have the same temperature to a few parts in $10^{5}$ over regions which were causally disconnected (outside of each other's Hubble radius) at the time of recombination $t_{rec}$ when the microwave photons last scattered. In standard big bang cosmology, the Hubble radius is (up to irrelevant factors of order 1) equal to the causal horizon, the maximal region of causal contact starting at the big bang. The seminal point of the horizon problem, therefore, is that in the context of standard cosmology there are no  causal microphysical processes which can explain the observed degree of isotropy which corresponds to correlations on these length scales.   

  A major success of inflationary cosmology \cite{Guth} is that it provides a simple solution of the Horizon problem. Inflation uses a  scalar (inflaton) field minimally coupled to gravity whose equation of state is assumed to be initially dominated by an almost constant potential energy and which consequently leads to exponential expansion of the Universe. Since in this case the causal horizon expands exponentially relative to the Hubble radius, inflation solves the horizon problem. Current inflationary models, however, generally are plagued by important problems of principle (see e.g. \cite{rhb99} for a recent review), for example fine-tuning of the potential required to obtain inflation and to obtain the desired magnitude of density fluctuations. Furthermore, Vachaspati and Trodden have recently argued \cite{VT98} that inflation simply shifts the causality problem to the level of having to demand acausally correlated initial data in order to obtain inflation (see, however, \cite{Kung} for a different point of view on this issue). In light of this, it is reasonable to investigate other possible solutions of the horizon problem.
  
	Recently Albrecht and Magueijo \cite{AM98} (see also Moffat \cite{Moffat} for earlier work) have proposed an alternative solution of the horizon problem by postulating a varying the speed of light (which determines the angle of the light cone).  The scenario is quite simple: If at early times the speed of light is much larger than at the present time, then the forward light cone at $t_{rec}$ will be much larger than it would be in standard cosmology, i.e. if computed with constant speed of light. One example of this scenario is if at a critical time $t_c < t_{rec}$ in the Universe's past, a ``phase transition'' occurs when the speed of light changes from $c^{-}$ to $c^{+}$, where $c^{-}$ is much larger than the current speed of light $c^{+}$.  

	 Kiritsis \cite{Kiritsis} recently discovered a realization of the varying speed of light scenario in the context of string/M theory \cite{Kiritsis2,Kraus}. He demonstrated that a 3 + 1 dimensional brane-world which probes a black hole background of stacked 3-branes obtains a speed of light which varies as the distance between the probe and the black hole changes. Since the system corresponds to a vacuum configuration of D-branes, residing in the Coulomb branch of string/M theory, this yields a rather natural realization of the varying speed of light scenario. In this note we will generalize Kiritsis's approach of solving the horizon problem with a varying speed of light, by proposing a mechanism which can stabilize the probe universe relative to the black hole and can thus lead to a speed of light which is constant at late times. We also address some other issues relevant to cosmology which inevitably arise in this framework.

\section{The Brane/Yang-Mills Setup}

	How is one to realize a varying speed of light in the context of string theory?  Since our visible universe occupies 3 spatial dimensions and string/M theory requires at least 11 for consistency, it is reasonable to look at vacuum configurations (points in moduli space) that predict non-trivial $3+1 D$ background space-times consistently.  Since D-branes are embedded hypersurfaces in the higher dimensional target space of all consistent string theories, it is reasonable to visualize our universe as being a D3 brane immersed in a higher dimensional bulk space-time. Recently, this brane-world interpretation has been utilized to solve the hierarchy problem \cite{RS}.  Likewise, this point of view motivates the realization of a variable speed of light in the context of string/M theory \cite{Kiritsis}.
 
One of the fascinating properties of D-branes is their ability to be stacked on top of each other without any cost in energy.  This is possible because they are BPS objects \cite{BPS}. When D-p branes with the same Ramond-Ramond charge are placed close to one another they experience zero net force, since their attractive gravitational force exactly cancels their electromagnetic repulsion.  Consequently, many branes with the same charge can be stacked coincident to each other.  For example, a stack of N D3 branes, defined by a $SU(N)$ gauge theory on its world volume, describes a point in the moduli space in string theory \cite{moduli}. 

 Furthermore, moving one brane a distance r away from the stack of N D3-branes (keeping the branes parallel) is in the corresponding gauge theory equivalent to turning on a Higgs expectation value $r/\alpha = <\phi>$ which leads to the gauge symmetry breaking pattern $SU(N) \rightarrow SU(N-1)\times U(1)$ \cite{moduli}.  Giving an expectation value to the scalar field associated with the position of the transverse position of the brane moves one onto the Coulomb branch \cite{Kiritsis} - a space of maximally supersymmetric but nonconformal vacua.  $\cal{N}= \rm 4$ $SU(N)$ super Yang-Mills theory includes the scalar fields $\Phi_{i} (i=1...6)$ which transform in the vector representation of the $SO(6)$ R symmetry group and in the adjoint representation of $SU(N)$.  The Coulomb branch corresponds to giving these fields expectation values subject to the flatness conditions of the potential $(\Phi)=[\Phi_{i},\Phi_{j}]=0.$  Hence these fields can be diagonalized and the moduli space is parametrized by the 6N eigenvalues $y_{i}^{(a)} (a=1...N)$ Since some scalars obtain non-vanishing expectation values in the Coulomb branch, the remaining massless bosons will mediate the long-range Coulomb interactions.

On the other hand, it is well understood through the AdS/CFT duality \cite{Maldacena}, that in the limit $N \rightarrow \infty$ and large `t Hooft coupling $\lambda$, the D brane stack is well described by supergravity with a specific background field configuration which is asymptotically $AdS$ space-time with an AdS black hole at the center.  One can study a probe D3 brane separated by a distance r from the AdS black hole 
with the $U(1)$ part of thermal $\cal{N} \rm = 4$ SYM $SU(N) \rightarrow SU(N-1)\times U(1)$.  
Based on this setup, Kiritstis \cite{Kiritsis} investigated the situation in which a probe D3 brane is moving in the vicinity of the horizon of  the $AdS_{5} \times S^{5}$ black hole, a black hole at finite temperature.  

   Since $N$ is large we can trust the supergravity solution and likewise treat the probe brane correctly with a Dirac-Born-Infeld (DBI) action.  Hence, one can interpret this process as our early universe emerging from the probe brane, while the higher dimensional black hole is fully described by the bulk supergravity.  The essence of Kiritsis'  calculation is to show how the fields living on the brane probe world volume (our Universe) are affected by the nontrivial higher dimensional bulk space-time (the blackhole) in which the brane is moving.  What we will see emerging is a variable speed of light on the probe brane world volume as the brane falls towards or away from the black hole.  This scenario raises some important issues that will be discussed in this note.

\section{Realization of The Varying Speed of Light Scenario}

	Given that we are going to do our analysis in the Coulomb branch, it is useful to exploit some well known technology.  In particular, we will be able to study the brane probe in the background of other branes by making use of the Dirac-Born-Infeld (DBI) action which describes the physics on the probe brane in the background of the 10D SUGRA solution which describes the collection of parallel stacked branes. The  DBI action involves the pullback of the bulk spacetime metric onto the brane world volume to determine the metric and necessary gauge fields localized on the brane.  Therefore, there exists a direct way to study the action of the bulk on both the fields and the space-time of the probe 3 brane universe.  

First, we will review  Kiritsis' way of obtaining a varying speed of light on the probe brane.  We begin with the relevant bosonic part of the action of Type IIB string theory:
\be 
S_{IIB}=\int d^{10}x\sqrt{-g}([R + 4(\partial\phi)^{2}] - \frac{2}{(8-p)!} F^2) 
\ee
where F is a RR-form field strength.  An object extended along p=3 spatial dimensions couples electrically to a 4 form gauge potential, and thus corresponds to a non-zero 5 form field strength.  As mentioned in the previous section, the consistent bulk spacetime solution within the Coloumb branch is the product manifold of $S^5$ and a five-dimensional AdS black hole, which corresponds to a p-brane solution.  In this case, the flux of the five-form field strength through $S^5$ carries quantized D3 brane charge.  The metric of the background geometry (in the string frame) of a near-extremal black hole describing N coinciding D3 branes is \cite{Kiritsis}:

\be 
ds^{2}= \frac{-f(r)dt^{2} + \vec{dx}^{2}}{\sqrt{H_{3}(r)}} + \sqrt{H_{3}(r)} (\frac{dr^{2}}{f(r)} + r^{2}d \Omega^{2}_{5}) 
\ee

where $$H_{3}(r)= 1+\frac{L^{4}}{r^{4}}$$
$$f(r) =1 - \frac{r_{0}^{4}}{r^{4}} \, . $$

The parameters $L$ and $r_{0}$ determine the $AdS$ throat size and the position of the Schwarzschild horizon, respectively and are both related to the ADM mass M and the integer Ramond-Ramond charge N.  We will deal with the spacetime in the near horizon $AdS$ geometry of the extremal $r_{0} \rightarrow 0$ black $D3$ brane solution.  Therefore, in the near horizon limit, $H(r) \rightarrow \frac{L^{4}}{r^{4}}$ and $L \simeq R$.  The resulting metric describes an $AdS-Schwarzschild$ black hole \cite{Hawking}:

\be 
ds^{2}= \frac{r^{2}}{R^{2}}(-f(r)dt^{2} + dx_{1}^{2} + dx_{2}^{2} + dx_{3}^{2}) + \frac{ R^{2} dr^{2} } {f(r) r^{2} } + R^{2}\Omega^{2}_{5} 
\ee

\noindent Here the 3 brane fills the coordinates $x_0 ... x_3$, $r$ is the direction transverse to the branes and $R$ is the radius of the $AdS_{5}$ spacetime and the five-sphere. 

	In the near horizon regime, one can take a probe brane far (but not so far that the near-horizon limit of the bulk metric breaks down) away from the black hole and evaluate the effects of the bulk space-time by using the DBI action. The probe is taken parallel to the stack of N three-branes to preserve supersymmetry.
%%?? is broken. Why do you have this? 
Hence, the DBI action is: 
  
\begin{eqnarray} 
S_{p} &=& -T_{p}\int d \tau d ^{p}\sigma e^{- \phi}\sqrt{-det(g_{\alpha \beta} \partial_{\mu} X^{\alpha} \partial_{\nu} X^{\beta} + F_{\mu\nu} - B_{\mu \nu}}) \nonumber \\  &+& T_{p}\int C^{p+1}
\end{eqnarray}

\noindent where $T_{p}$ is the Dp-brane tension, the R-R charge is the source to the p-form field $C_{p}$, $F$ and $B$ are the gauge fields living on the brane world volume,  $\phi(x)$ is the dilaton field and $X^{\alpha}$ are the target space coordinates.
   
	Given that a brane probe is often embedded in a curved bulk space-time, the pullback of such a background onto the brane world volume needs to be computed directly.  The existence of the curved background and the fact that the induced metric on the probe brane depends on the background fields are the main reasons why the  speed of light on the brane can vary in time in this scenario.  Hence, this calculation will make the origin of the varying speed of light on the probe brane transparent. 

First, let us expand the action keeping quadratic terms in derivatives .

Here G is the determinant of the worldvolume metric $G_{\mu \nu}= g_{\alpha \beta} \partial_{\mu} X^{\alpha} \partial_{\nu} X^{\beta}$. Since we know $ 
g_{\alpha\beta} $, the near horizon black hole metric, we are able to find the induced world volume metric.  We use the static gauge 
$$X^{0}=t$$
$$X^{\mu}=x^{\mu}$$
$$X^{4}=r$$ $$X^{\gamma}=x(\Omega_{5}),$$  
where $\mu=1...3$ and $\Omega_{5}$ represent the coordinates on the five sphere. 

Since the gauge fields are smaller in magnitude than the curvature, i.e. $|F| < |G|$, we can expand the DBI action using the identity:
\be 
det^{1/2}(1 + M)=exp( \frac{1}{2} tr[M - \frac{1}{2} M^{2} + \frac{1}{3} M^{3} + . . .] ) 
\ee
with $M^{\mu}_{\nu}=F_{\mu \sigma} G^{\sigma \nu} $

Now we are to look at the first order kinetic terms of the gauge fields living in the brane.  For simplicity, we set $B_{\mu\nu}=0$:

\begin{eqnarray} 
S_{GF} &=& T_{3} \int d ^{4}x \frac{1}{\sqrt {f(r)}} \large{[} G^{\mu \rho} G^{\nu\sigma} F_{\mu\nu}F_{\rho\sigma} \large{]}\nonumber \\ 
&=& 2 T_{3} \int d^{4} x \large{[} \frac{1}{\sqrt {f(r)}}\vec{ E}^{2} + \sqrt {f(r)} \vec{B}^{2} \large{]} \, .
\end{eqnarray}

\noindent where $E_{i}=F_{0i}, B_{i}=\epsilon_{ijk} F_{jk}/2.$ 

\noindent Here we see that the function f(r) that appears in front of the electric and magnetic fields, respectively, is indeed the speed of light that appears in the Maxwell equations:

\be 
c_{eff}=\sqrt{f(r)}=\sqrt{1-(\frac{r_0}{r})^{4}} \, . 
\ee

\noindent From this equation we see that the speed of light decreases as the probe-brane universe approaches the black hole and vanishes at the horizon.  Hence, from this analysis we are able to realize a varying speed of light in the context of a brane universe.  However, our Universe today does not have a vanishing speed of light.  It is therefore important to investigate how the speed of light can be stabilized in the context of the Coulomb branch picture.  In the next section we will build on Kiritsis' analysis by generalizing the dynamics of the system to include rotation.

\section{Inclusion of Rotation}

	Given that the speed of light can vary on a probe 3-brane universe as it falls into or moves away from a black hole, it is natural to investigate how this situation will be affected by generalizing the scenario to a rotating black hole. In other words, what will happen to the speed of light on the brane if it falls into a rotating black hole with a definite angular momentum?  Also, how can we stabilize the probe brane in some fixed orbit, thus, rendering the speed of light constant at late times? In order to address these questions, it is important that the analysis remain consistent with the Coulomb branch analysis of Kiritsis. Thus, we aim to study the same system of a brane probing a black hole and generalize it to include rotation.  Interestingly, it has been demonstrated \cite{Cai} that a rotating collection of D3 branes will still remain in the Coloumb branch, and in cases relevant to our analysis, maintain thermodynamic stability. Therefore, one can still trust the probe brane scenario.
  
	We begin with a collection of N branes which are rotating.  Following Cai \cite{Cai}, we consider the rotating D3-brane solution with an angular momentum parameter. The metric is

\begin{eqnarray}
ds^{2} &=& \frac{1}{\sqrt{H}} (-fdt^{2} + dx^{2}_{1} + dx^{2}_{2} + dx^{2}_{3}) 
\nonumber \\
&+& \sqrt{H} \large{[} \frac{dr^{2}}{\tilde{f}} - 
\frac{4mlcosh \alpha}{r^{4} \Delta H} sin^{2}\theta dt d \phi \nonumber \\ 
&+& r^{2}(\Delta d\theta^{2} + \tilde{\Delta} sin^{2} \theta d \phi^{2} + cos^{2} \theta d \Omega^{2}_{3} \large{]} 
\end{eqnarray}

\noindent where 
\be 
H=1 + \frac{2msinh^{2} \alpha}{r^{4} \Delta}, 
\ee
\be 
\Delta = 1 + \frac{l^{2} cos^{2}\theta}{r^{2}}, 
\tilde {\Delta}= 1 + \frac{l^{2}}{r^{2}} + \frac{ 2m l^{2} sin^{2} \theta }{r^{6} \Delta H} 
\ee
\be 
f= 1- \frac{2m}{r^{4} \Delta}, \tilde{f}= \frac{1}{\Delta} \large{[} 1 + \frac{l^{2}}{r^{2}} - \frac{2m}{r^{4}} \large{]}. 
\ee

Here $\phi$ is the angle in the plane in which the black hole is spinning and the angles in the 3-sphere specify a direction labeled $rcos \theta$ above and orthogonal to this plane.  The remaining angle $\theta$ runs between the plane and the the $rcos \theta$ direction.  The term $sinh\alpha$ is proportional to the total D3 brane charge $N_{D3}$. 
The rotating D3 brane is supported by the self-dual 5-form field strength $F_{5}$ of Type IIB string theory.

Note that the action for the varying speed of light of the gauge fields, eq (6),  will carry over to the case of the rotating background.  Hence, there will be no angular momentum dependence of the speed of light on the probe brane to first order in its velocity.

Using the DBI action (4) one obtains the low velocity action of a D3 probe brane moving in a rotating background of N D3-branes:

\begin{eqnarray}
S &=& -T_{3} V_{3} \int d \tau H^{-1} [ \sqrt{f-H w^{2}} - 1 + H_{0} 
\nonumber \\
&-& H + \frac{ (1-H) lsin^{2} \theta }{sinh \alpha } \dot{\phi} ]  
\label{act} \end{eqnarray}

\noindent where 

\begin{eqnarray} 
w^{2} &=& \frac{ \dot r^{2} }{ \tilde{f} } + r^{2} (\Delta \dot{\theta}^2 + \tilde{\Delta} sin^{2}\theta \dot{\phi^{2}} + cos^{2} \theta \dot{\Omega^{2}_{3}}) \nonumber \\
&-& \frac{4mlcosh \alpha}{r^{4} \Delta H} sin^{2} \theta \dot{\phi} .
 \end{eqnarray}

\noindent and 
\be 
H_{0} =1 + \frac{ R^{4} }{ r^{4 } \Delta }. 
\ee

This action generalizes the probe brane action to a rotating background.  Minimizing action \ref{act} yields the geodesic equations of motion of the probe brane universe in the rotating black hole background.  Since the speed of light is stabilized by keeping the probe brane at a fixed radial distance from the black hole, $c(r_{const})_{eff}= const$, we need to find a fixed orbital motion from the resulting geodesic equations of motion \ref{act}.  This procedure is similar to finding circular geodesics for Schwarschild black holes.  

We vary action \ref{act} with respect to $r$ and the angle in the plane of rotation $\phi$.  We will look at a special case of solutions, whereby the brane-universe is rotating perpendicluar to the plane of rotation of the black hole.  This configuration corresponds to setting $\theta=\frac{\pi}{2}$.  The resulting equation of motion for variation with respect to $r$ is:

\be   K \dot{\phi}= H^{-1}_{,r}(W)^{1/2} + \frac{1}{2}H^{-1} W^{1/2}_{,r} + \tilde{H_{,r}} \label{eomr} \ee
The second equation of motion with respect to the angular variation, $\phi$ is:

\be 2r^{2}\ddot{\phi}\dot{\phi} - \frac{1}{2}W^{-1/2}[\gamma\ddot{\phi} -2H\ddot{\phi} \dot{\phi}](2r^{2} \dot{\phi} - \gamma) = 0 \label{eomphi} \ee
where
\be K=\frac{l\sin{\theta}}{\sinh{\alpha}} \ee
\be \tilde{H} = H + H_{0} \ee
\be W = (f-Hw^{2}) \ee
\be \gamma =\frac{4mlcosh \alpha}{r^{4} \Delta H} sin^{2}\theta \ee

In order for the brane-universe to have a stable circular geodesic around the black hole, the equations of motion \ref{eomr} and \ref{eomphi} have to simultaneously satisfy the following conditions:
\be \ddot{r}=\dot{r}=o \ee
This corresponds to a fixed radial distance from the black hole horizon.

\be \phi=\omega t \ee

This condition corresponds to a periodic angular motion with a constant angular frequency, $\omega$.
Therefore, circular orbits require this specific behavior of the radius and the angular velocity.  We now show that the geodesic E.O.M are solved for circular orbits.

After some algebra one sees that the R.H.S of equation \ref{eomr} is a constant; it immediately follows that this equation satisfies the condition for $\phi=\omega t$, since the left hand side is proportional to $\dot{\phi}$, which is also constant.  Likewise one sees that the second eqaution \ref{eomphi} is satisfied since $\ddot{\phi}=0$, hence the left hand side is equal to zero with the solution for a circular geodesic. In other words, the angular momentum of our system, $\frac{\partial{L}}{\partial{\phi}}= p_{\phi}$, is a constant of the motion. We therefore conclude that the geodesic equation of motion for the brane-universe rotating around the black-hole admits circular motion.

 Since the speed of light on the brane vanishes at the horizon, we have to stabilize the brane in an orbit away from the horizon.  It is well known that rotation pulls the brane in the direction of the local inertial frame. This is a frame dragging effect which will deflect the falling brane from the ergosphere which is situated outside the horizon.  Hence, we see that including rotation keeps the brane-universe from falling into the horizon.  This step is crucial, but not sufficient to stabilize the brane at a fixed radius away from the horizon, and thus not sufficient to provide a theory in which the speed of light stabilizes at a small value at late times.

We wish to propose a natural mechanism for stabilizing the speed of light on the probe-brane at late times to a small value. Crucial ingredients are rotation of the black hole (as mentioned above), expansion of the bulk space-time, and virialization of bulk density fluctuations. If the bulk contains matter, the bulk will expand. This expansion leads to a decrease in peculiar velocities of probe particles and probe branes. Thus, a brane which approaches the black hole and is deflected by the rotation will not move away at the same speed.  Rather, it will lose energy and gradually spiral towards the black hole.  Since the expansion rate of the bulk will decrease as time increases (as our Universe's expansion rate is decreasing as time increases), the damping effect will weaken.

 In addition, the region of bulk space-time surrounding the black hole corresponds to a bulk density fluctuation.  This fluctuation will increase in magnitude by the usual gravitational instability mechanism and may freeze out from the general Hubble expansion once the density contrast reaches order unity, analogous to how galaxies freeze out from the universal expansion of our Universe once the density contrast in the galaxy reaches order unity. Afterwards, the region of bulk space-time in the vicinity of the black hole will virialize like a galaxy virializes \cite{Peebles} and settles into a stable angular motion in a four-dimensional expanding Universe. After freezeout and virialization, the probe particle peculiar velocities will no longer decrease. Therefore, the probe brane will settle into stable rotation about the black hole, thus achieving a state with constant (and small) speed of light.

\section {Discussion and Conclusion}

	In this letter we have proposed a realization of the varying speed of light scenario for solving the horizon problem in the context of string/M theory. It is based on an extension of the analysis of Kiritsis \cite{Kiritsis} in which the observed four dimensional Universe is a probe brane in a higher dimensional bulk space-time in which a black hole produced by a stack of 3 branes. As Kiritsis realized, the speed of light on the probe brane decreases in time as the probe brane moves towards the black hole in the bulk space-time. The new ingredients in our approach are the use of a rotating black-hole background to prevent the probe brane from hitting the horizon, the use of bulk expansion to provide an effective friction for the probe brane dynamics, which
enables the brane to remain close to the black hole at late times; and the freezeout of the bulk space-time near the black hole, and subsequent virialization to stabilize the radius of the probe brane orbit.  This will ensure that the speed of light tends to a constant at late times.

	Related to this work is the following question: Can one describe the dynamical capture of the probe brane quantum mechanically by an absorption cross section? This has been successfully realized in the context of scalar field absorption using the AdS/CFT correspondence\cite{Klebanov,Das}. We are currently investigating the rotating black hole background required for our scenario and will report on our findings in future work.

\acknowledgements
I wish to thank Robert Brandenberger and Sanjaye Ramgoolam for the illuminating discussions and encouragement throughout the course of this project.  Also, I thank Joao Magueijo for inspiring me to write this paper.  I thank the GAANN fellowship program for the financial support.

\end{document}